\documentclass[sigconf]{acmart}
\usepackage[caption=false]{subfig}
\AtBeginDocument{%
  \providecommand\BibTeX{{%
    \normalfont B\kern-0.5em{\scshape i\kern-0.25em b}\kern-0.8em\TeX}}}

\usepackage{todonotes}
\usepackage{enumitem}

\setcopyright{acmcopyright}
\copyrightyear{2022}
\acmYear{2022}

\acmConference[CHI '22]{Conference on Human Factors in Computing}{April 29-May 5, 2022}{New Orleans, LA, USA}
\acmBooktitle{Conference on Human Factors in Computing Systems, April 29-May 5, 2022,
New Orleans, LA, USA}
\acmPrice{15.00}
\acmISBN{978-1-4503-XXXX-X/18/06}

\begin{document}

\title{Supporting Electronics Learning through Augmented Reality}


\author{Thomas Kosch}
\email{thomas.kosch@hu-berlin.de}
\affiliation{%
  \institution{HU Berlin}
  \city{Berlin}
  \country{Germany}
}

\author{Julian Rasch}
\email{julian.rasch@ifi.lmu.de}
\affiliation{%
  \institution{LMU Munich}
  \city{Munich}
  \country{Germany}
}

\author{Albrecht Schmidt}
\email{albrecht.schmidt@ifi.lmu.de}
\affiliation{%
  \institution{LMU Munich}
  \city{Munich}
  \country{Germany}
}

\author{Sebastian Feger}
\email{sebastian.feger@ifi.lmu.de}
\affiliation{%
  \institution{LMU Munich}
  \city{Munich}
  \country{Germany}
}

\renewcommand{\shortauthors}{S. Feger et al.}

\begin{abstract}
Understanding electronics is a critical area in the maker scene. Many of the makers' projects require electronics knowledge to connect microcontrollers with sensors and actuators. Yet, learning electronics is challenging, as internal component processes remain invisible, and students often fear personal harm or component damage. Augmented Reality (AR) applications are developed to support electronics learning and visualize complex processes. This paper reflects on related work around AR and electronics that characterize open research challenges around the four characteristics functionality, fidelity, feedback type, and interactivity.
\end{abstract}



\begin{teaserfigure}
  \includegraphics[width=\textwidth]{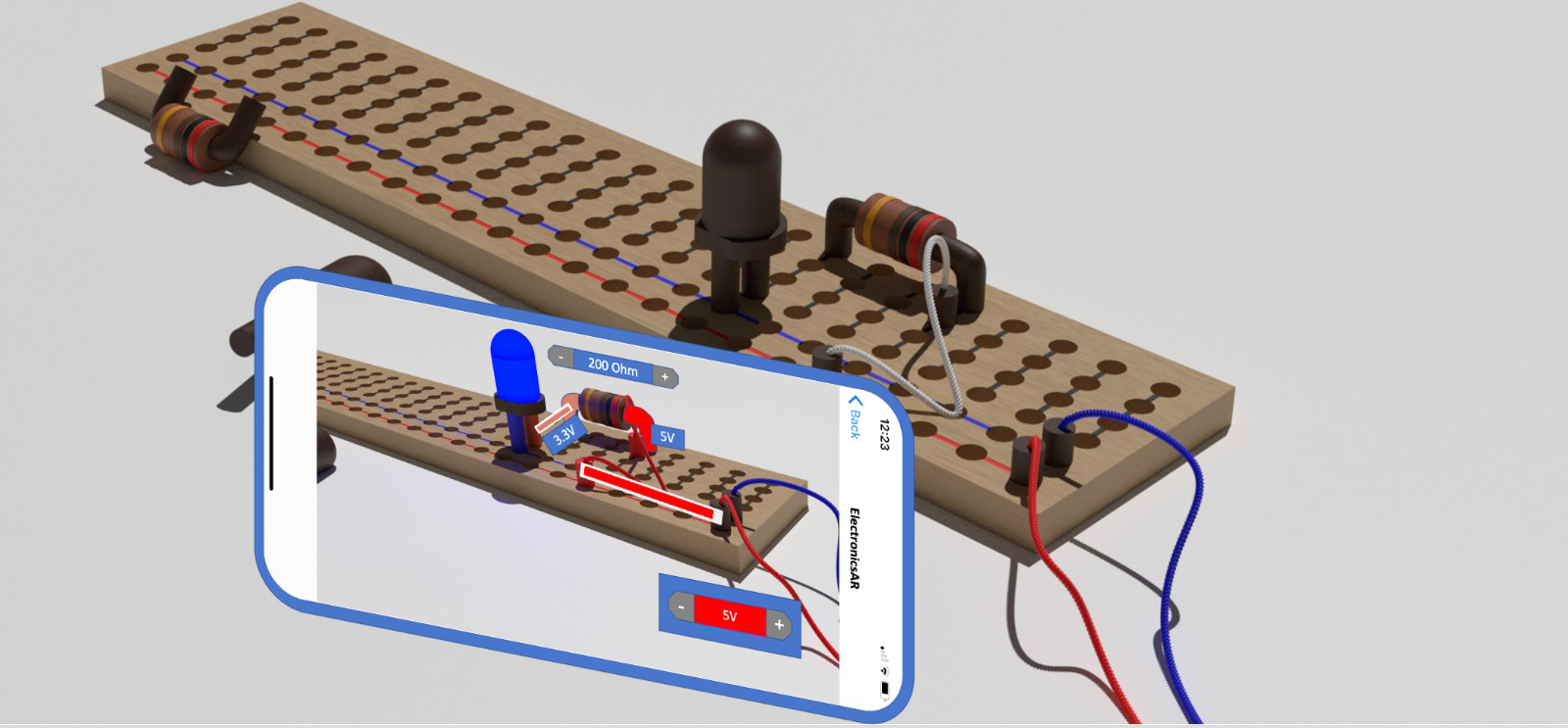}
  \caption{A mock up of a non-functional augmented electronics breadboard. A smartphone app analyzes the functionality and correctness of the circuit by detecting the configured circuit and spatially projecting virtual elements on the circuit components.}
  \label{fig:teaser}
\end{teaserfigure}

\maketitle

\section{Introduction and Background}
Electronics is an essential part of STEM\footnote{Science, Technology, Engineering and Mathematics} education. Experimenting with electronics is subject to a series of unique challenges for beginners, and advanced learners \cite{menendez2006virtual, ferreira2014enriched}. Yet, the inability to perceive the internal processes of electronic components and circuits is a significant barrier. Consequently, educators and technology developers explored analogies, such as ski lifts and bowling balls \cite{mogstad2020ski}, as well as water pipes and bike chains displayed to explain electronics and circuits. Although rich analogies that convey the basic principles of electrical flow exist, they do not practically teach how single components affect and are affected within an electrical circuit. Consequently, concerns about injuries, circuit complexity, and fear of damaging hardware are barriers for learners who delve into basic electronics. Research explored how interactive systems can facilitate and teach electronics to foster hands-on creativity. We position previous research projects that augment the user's understanding of electronic circuits, components, and output in this work. We describe the projects and discuss open research challenges for the maker community.


\section{Technology-Mediated Electronics Support}
The inability to visualize internal circuit processes, the fear of injury, and concerns to damage hardware, are significant key barriers of education in basic electronics \cite{dobson1995evaluation, somanath2017maker}. The development of technology-mediated self-directed learning strategies to these problems is as diverse as the challenges they aim to address. This wide range of interactive systems designed to support electronics users in training and normal circuit building range from AccessibleCircuits \cite{accessibleCircuits}, providing 3D-printed add-on circuit components to help persons with visual impairments in constructing circuits to smart functional breadboards that detect, digitize, and analyze configurations \cite{toastboard, currentviz, circuitsense}.

AutoFritz \cite{autofritz} is an example of a completely screen-based support system that helps users design circuits on a popular prototyping platform, Fritzing, through suggestions and autocompletion. AutoFritz provides recommendations based on an analysis of corresponding datasheets and a comprehensive set of openly available community projects. In contrast, Proxino \cite{proxino} is an example of a tool that bridges between entirely virtual screen-based circuits and physical components. Proxino enables interaction with virtual circuits through physical proxies and supports remote collaboration and learning. Similarly, Simpoint \cite{simpoint} makes use of a tight coupling between real physical circuits and virtual counterparts. Here, Strasnick et al. created a system that juxtaposes live signals measured from a physical circuit with simulated data from this circuit's model to debug. Simpoint further allows modifying the signals and component parameters as advanced debugging features. Previous research showed that blending real and physical information supports learners when interacting with physical circuits.

On the other end of the spectrum, systems have been developed that focus on the augmentation, analysis, and interaction with the physical electronics components. For example, Drew et al. \cite{toastboard} developed the Toastboard, a type of extended smart breadboard that measures the voltage of each row. An LED bar directly indicates one of three types of voltages detected: power, ground, or other voltage, enabling users to perform the first set of analyses on the breadboard. Those measurements can be shared with a dedicated software application. SchemaBord \cite{schemabord} by Kim et al. makes use of LEDs to provide additional information. SchemaBoard is an LED-backlit functional breadboard that can highlight elements selected in a circuit schematic displayed on a connected computer. This augmented breadboard is expected to support makers in finding, placing, and debugging their real circuits. Another example is CurrentViz \cite{currentviz}, a system similar to Toastboard but measuring current instead of voltage. Finally, CircuitSense \cite{circuitsense} is a smart breadboard that detects the locations of placed components and recognizes the component types automatically. This way, users can quickly create virtual circuits by digitizing their physical counterparts. This is expected to benefit the open sharing of circuit designs. Overall, previous research demonstrates the value of augmented physical breadboards as primary interaction material. Here, augmentations were limited to LED bars and computer-processed visual information.

\begin{figure*}
    \centering
    \centering
    \subfloat[][]{
        \includegraphics[height=0.39\columnwidth]{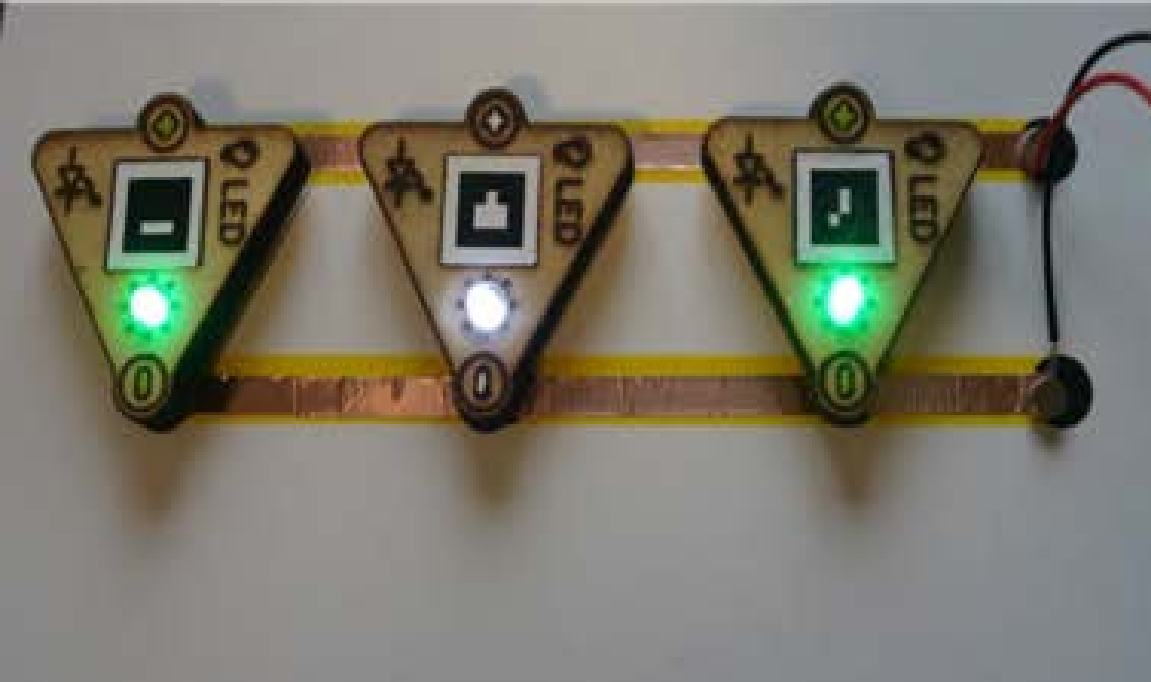}
        \label{fig:arbits}}
    \subfloat[][]{
        \includegraphics[height=0.39\columnwidth]{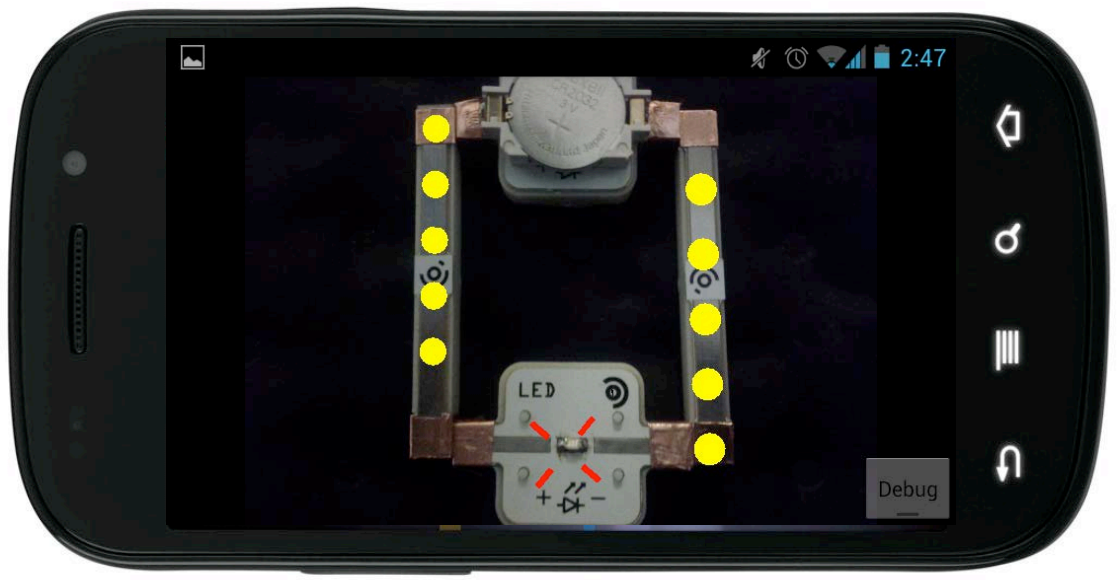}
        \label{fig:lightup}}
    \subfloat[][]{
        \includegraphics[height=0.39\columnwidth]{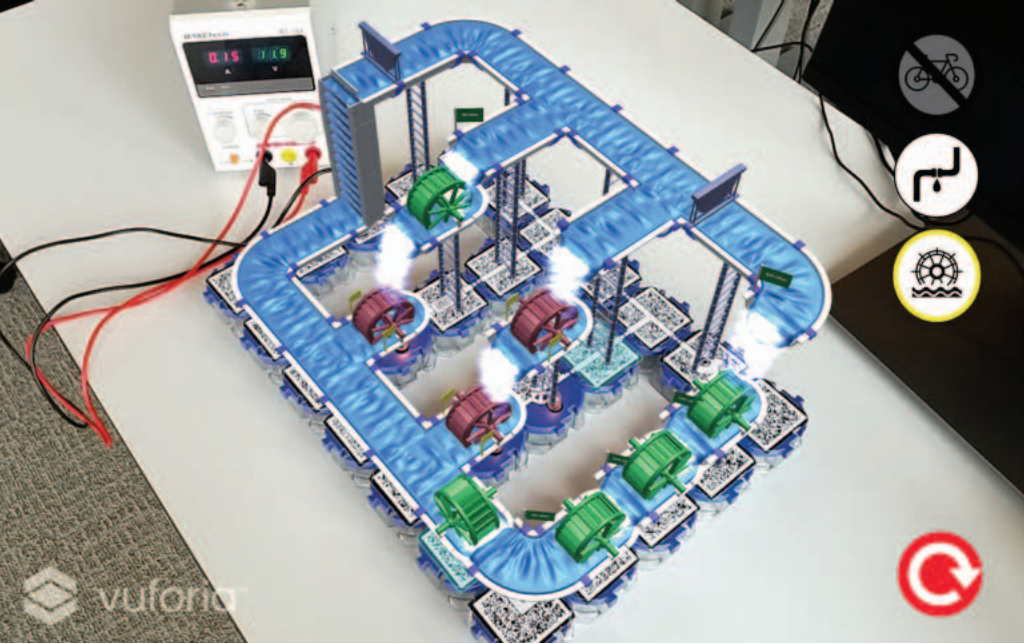}
        \label{fig:arcc}}
    \caption{Augmentation strategies for electrical circuits. \textbf{(a):} ARbits uses laser cut tangibles for circuits and embeds the component functionality into tangibles~\cite{arbits}. \textbf{(b):} LightUp uses a smartphone app to analyze circuits and show information about the electron flow~\cite{lightUp}. \textbf{(c):} The AR Circuit Constructor uses the waterfall analogy to mediate the principles of electrical flow~\cite{ARCircuitConstructor}.}
    \label{fig:arbits-lightup-arcc}
\end{figure*}

\section{Augmenting Circuits}
Several AR applications were designed to support users in understanding the internal processes of circuits and electronic components. For example, ARBits \cite{arbits} is a toolkit consisting of significant wooden parts, where each component represents functional components (see Figure~\ref{fig:arbits}). The kit comes with wooden blocks and operational features for eight-element types, including LEDs, DC motors, motor drivers, buzzers, object detection sensors, potentiometers, LED matrices, microcontrollers. Each wooden block further features one printed AR marker used for object detection and recognition. The blocks are more significant than the components they represent. This makes them suitable for interaction with students at an early age. Yet, their shape is abstract and barely related to the shape of the components represented. Aligned in a circuit, the electronic components are fully functional. At the same time, the users can visualize basic information, including component type and polarity, through a mobile AR app.

Chan et al. \cite{lightUp} developed LightUp, an electronics kit similar to ARBits (see Figure~\ref{fig:lightup}). LightUp also makes use of functional but enlarged and abstract components. In addition, LightUp even provides functional bricks for the conductor paths and comes with a pre-mounted battery block. Yet, the visualizations provided through the mobile AR app are more limited and focus mainly on bubbles indicating the order in which the elements are configured within the circuit. In contrast, AR Circuits \cite{arCircuits} is limited to printed paper cards with AR trackers representing a type of electronic component. Students align those cards to create a non-functional circuit that can be visualized and analyzed through a mobile AR app. Notably, some of the displayed components, for example, the button, are interactive and impact the circuit behavior.

Kreienbühl et al. \cite{ARCircuitConstructor} designed AR Circuit Constructor (ARCC), a toolkit featuring abstract but functional electricity building blocks including essential components and a QR code used for AR tracking and identification (see Figure~\ref{fig:arcc}). In contrast to ARBits, LightUp, and AR Circuits, the mobile AR application of ARCC focuses on providing detailed circuit feedback. Students can choose between three types of analogy-driven visualizations: bicycle chains, water pipes, and waterfalls. The \textit{water pipe} visualization, for example, displays functional circuits as a closed water pipe system in which resistors are represented by lines that are tighter (i.e., the diameter of the pipe reflects the resistance). In addition, the authors conducted a qualitative user study with eight science teachers and found that educators would use ARCC for self-directed explorative learning \cite{roussou2004learning}.

Reyes-Aviles and Aviles-Cruz \cite{reyes2018handheld} proposed a system that differs strongly from previously presented approaches in terms of the level of feedback provided. Their mobile app expects to receive a photo of a functional breadboard and a resistor circuit configuration, along with captured data about the voltages and currents measured at the nodes of installed resistors. The application performs image recognition for all resistors installed on the breadboard and adds information related to the calculated theoretical voltage and power consumption within each resistor node directly into this augmented image. This approach has inspired further work in augmenting electronics using augmented reality to foster learning~\cite{feger2022electronicsar}.

Kosch et al.~\cite{kosch2017chances} and Knierim et al.~\cite{knierim2018challenges} investigated how mixed reality can be used to teach the construction of basic electronic circuits. Information about the components and instructions to construct an electrical circuit was projected on a table. Objects were detected using a depth sensor and computer vision. The learners can load an electronics exercise and are guided step-by-step through the construction process. Here, students learn the basic principles of the parts, how they influence the circuit, and how they are assembled in a realistic scenario.

Wang and Cheung \cite{wang2021augmented} explored a similar research direction that relies on electronic component recognition in images. The authors expect their computer vision system to facilitate circuit building through step-by-step instructions in a mobile AR app. In contrast, Chatterjee et al. \cite{augmentedSilkscreen} explored how AR applications can be used to support experts in circuit debugging. They found that their proposed AR interaction technique, Augmented Silkscreen, can benefit experts in different ways, including searching for components and probe points on complex Printed Circuit Boards (PCBs) and element metadata. Yet, their evaluation has been limited to a PCB simulator and video sketches.

\section{Open Research Challenges}
The reflections in this section showed that students, learners, and experts are confronted with various challenges around tinkering with electronics, circuit building, and debugging. AR provides a solid opportunity to address those barriers by visualizing invisible and difficult processes to understand. This section also highlighted the diversity of approaches that can be characterized across a set of different features. This includes:

\begin{itemize}[leftmargin=*]
    \item \textbf{Functionality}. While some applications like AR Circuit Constructor \cite{ARCircuitConstructor} and ARBits \cite{arbits} make use of functional components and circuits, others focus on non-functional toolkits that are easy to set up and safe, or even entirely image-based. Examples include AR Circuits \cite{arCircuits} and the computer vision system presented by Wang and Cheung~\cite{wang2021augmented}.
    \item \textbf{Fidelity}. Related work used a wide range of approaches in terms of component fidelity. The spectrum ranges from simple paper-based trackers in AR Circuits \cite{arCircuits}, abstract blocks in ARBits \cite{arbits}, AR Circuit Constructor \cite{ARCircuitConstructor}, and LightUp \cite{lightUp}, to image-based augmentations of real circuits, as presented by Reyes-Aviles and Aviles-Cruz~\cite{reyes2018handheld}.
    \item \textbf{Feedback Type}. The feedback provided through AR applications ranges from abstract visualizations and in-situ projections~\cite{kosch2017chances} to detailed calculations. For example, LightUp \cite{lightUp} is limited to bubbles highlighting the overall circuit configuration. In contrast, AR Circuit Constructor \cite{ARCircuitConstructor} provides three analogy-driven visualizations that are expected to support students' understanding of voltage and current. At the other end of the spectrum, Reyes-Aviles and Aviles-Cruz \cite{reyes2018handheld} combine real measurements and component recognition to display accurate voltage and power usage. The type of feedback reflects the target audience: children, advanced students, or experts.
    \item \textbf{Interactivity} Most AR electronics apps presented in this section do not foresee user interaction with the virtual objects. Instead, almost all applications focus on displaying information. Exceptions include AR Circuits \cite{arCircuits}, allowing users, for example, to interact with virtual switches, thereby changing the state of the circuit.
\end{itemize}

\begin{figure}
    \centering
    \includegraphics[width=\columnwidth]{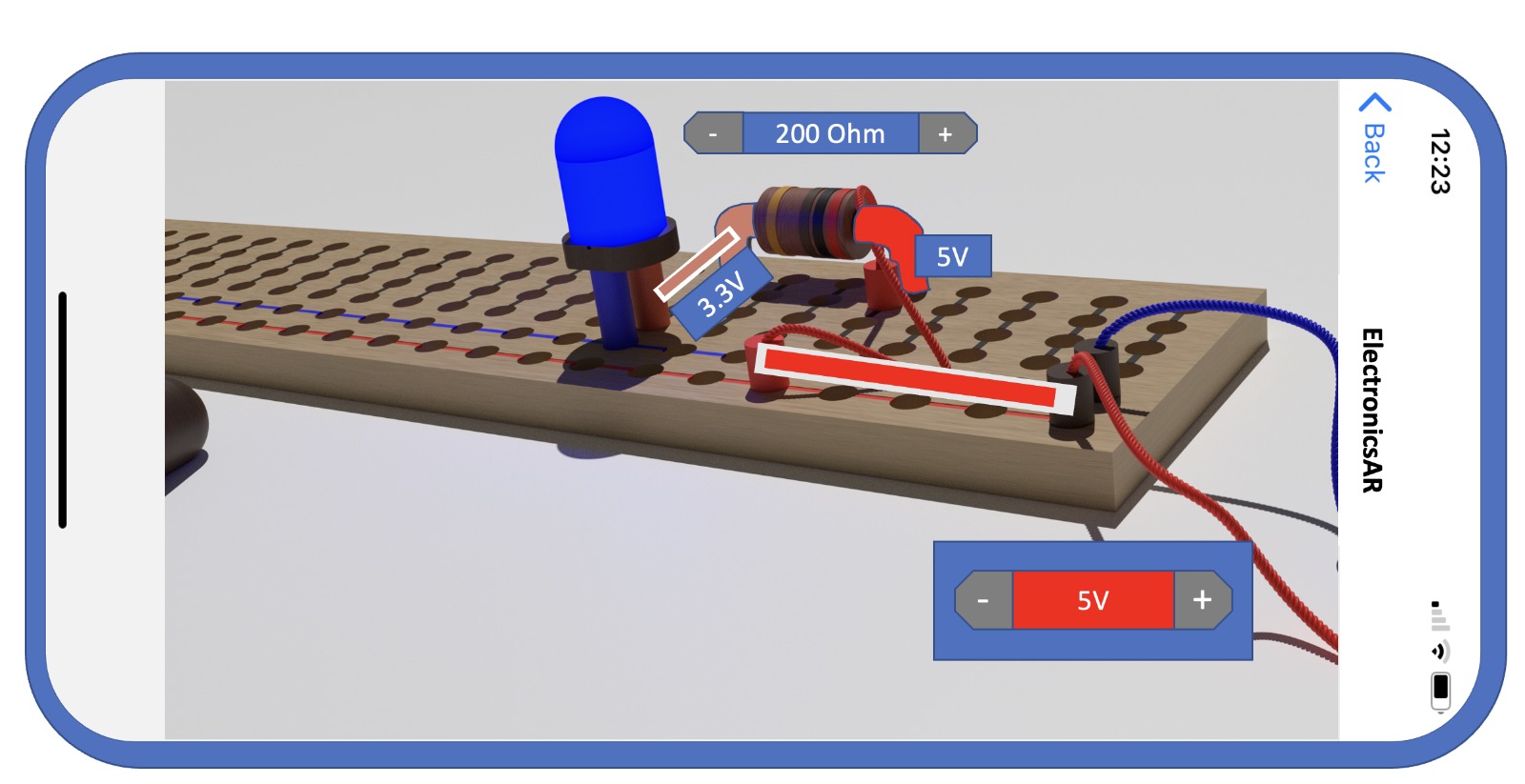}
    \caption{3D printed non-functional electronic parts viewed through an AR app. The envisioned toolkit detects the electrical components and their parameters, the electrical flow, and provides options to modify the parameters.}
    \label{fig:demo}
\end{figure}

\noindent While this summary does not aim to provide a systematic description of AR electronics characteristics; it helps identify the current challenges. Previous work showed how \textbf{non-functional} components are suitable for learners to understand basic electronics. However, representing electronics components in a \textbf{high-fidelity} remains a research challenge. Previous abstract representations \cite{arCircuits, arbits, lightUp, ARCircuitConstructor} do not communicate a high-fidelity metaphor for electronic components. Here, 3D printing is a viable alternative to rapidly create customized non-functional electronics parts in different sizes and levels of fidelity. The electronics components can be analyzed using augmented reality, where \textbf{detailed feedback} is provided (e.g., voltage or power). A sophisticated interaction design in AR enables learners to \textbf{interactively} engage and understand circuit designs. Users can set, change, and experiment with a wide range of values (see Figure~\ref{fig:demo}). We expect that this feature will help both young and advanced learners through self-directed explorative learning \cite{roussou2004learning}, as well as support advanced learners in understanding the effects of individual electronic parameters. However, sophisticated interaction designs can cause a novelty effect suggesting a subjectively perceived interaction~\cite{10.1145/3529225}. Hence, future research must investigate how placebo control conditions can be incorporated into evaluation studies.

\section{Conclusion}
We presented an overview of work in augmented electronics for circuit debugging and learning support. Based on this review, we presented current research challenges around the characteristics \textit{functionality}, \textit{fidelity}, \textit{feedback type}, and \textit{interactivity}, that we would be excited to discuss with the workshop participants. Further, we showcased our vision for future high-fidelity toolkits in Figures \ref{fig:teaser} and \ref{fig:demo}. We are interested in discussing how this vision can become a reality benefiting the wider maker space.

\bibliographystyle{ACM-Reference-Format}
\bibliography{sample-base}

\appendix

\end{document}